\documentstyle[aps]{revtex}

\def \kay {{p\over\rho}}
\def \kbar {{c_s^2}}
\def \cee {{C}}
\def \eor {{\left(1+\frac{p}{\rho}\right)}}    

\def \psir {{\hat\psi}}
\def \phir {{\hat\phi}}
\def \mur {{\hat\mu}}
\def \nur {{\hat\nu}}
\def \ur {{\hat{u}}}
\def \nr {{\hat{n}}}
\def \Ur {{\hat{U}}}
\def \Wr {{\hat{W}}}
\newcommand{\dr}[1]{\hat{\dot{#1}}}
\newcommand{\pr}[1]{#1^{\hat\prime}}

\begin{document}


\draft

\title{Gauge-invariant and coordinate-independent perturbations of
stellar collapse I: the interior}

\author{Carsten Gundlach} \address{Max-Planck-Institut f\"ur
Gravitationsphysik (Albert-Einstein-Institut), Schlaatzweg 1, 14473
Potsdam, Germany} \address{Enrico Fermi Institute, University of
Chicago, 5640 Ellis Avenue, Chicago, IL 60637 \footnote{Present
address}}

\author{Jos\'e M. Mart\'\i n-Garc\'\i a}
\address{Laboratorio de Astrof\'\i sica Espacial y F\'\i sica 
         Fundamental, Apartado~50727, 28080~Madrid, Spain}
\address{Universidad Alfonso X El Sabio, 
         Avda. de la Universidad 1,
         28691~Villanueva de la Ca\~nada, Madrid, Spain 
         \footnote{Present address}}

\date{16 June 99}

\maketitle


\begin{abstract}
  
Small non-spherical perturbations of a spherically symmetric but
time-dependent background spacetime can be used to model situations of
astrophysical interest, for example the production of gravitational
waves in a supernova explosion. We allow for perfect fluid matter with
an arbitrary equation of state $p=p(\rho,s)$, coupled to general
relativity. Applying a general framework proposed by Gerlach and
Sengupta, we obtain covariant field equations, in a 2+2 reduction of
the spacetime, for the background and a complete set of
gauge-invariant perturbations, and then scalarize them using the
natural frame provided by the fluid. Building on previous work by
Seidel, we identify a set of true perturbation degrees of freedom
admitting free initial data for the axial and for the $l\ge2$ polar
perturbations. The true degrees of freedom are evolved among
themselves by a set of coupled wave and transport equations, while the
remaining degrees of freedom can be obtained by quadratures. The polar
$l=0,1$ perturbations are discussed in the same framework. They
require gauge fixing and do not admit an unconstrained evolution
scheme.

\end{abstract}



\section{Introduction}

In many situations of astrophysical interest, spherical symmetry is a
good approximation for modeling a star in general relativity. One
possible direction in which to go beyond that approximation is to
allow for arbitrary linear perturbations, in order to add
gravitational radiation to the picture. This introduces new physics,
as the star can now lose energy through gravitational radiation, and
a new window of observation, as this gravitational radiation can be
detected. Gravitational wave detectors are expected to operate at the
necessary sensitivity for the first time within a few years, and a
large effort is under way to model possible sources of gravitational
radiation. If one allows the spherical background solution to be
time-dependent, as we shall do here, one can model for example the
gravitational radiation emitted in a (slightly nonspherical) supernova
explosion.

We assume here that the matter content is a perfect fluid described by
an equation of state $p=p(\rho,s)$ where $p$ is the pressure, $\rho$
the total energy density, and $s$ the entropy per particle. As a
consequence of the perfect fluid approximation, $s$ is assumed to be
constant along particle trajectories, that is, we neglect the possible
sources of entropy generation: heat fluxes, viscosity and chemical
reactions. We also assume that there is only a single fluid present.

Both the assumptions of approximate spherical symmetry and perfect
fluid matter may be unrealistic for supernovae. Some supernovae are
now conjectured to be quite nonspherical, and neutrino transport is
believed to play an important role. Here we concentrate on giving a
clean mathematical description of an almost spherical perfect fluid,
in the belief that this approximation will be useful in some
applications. 

There are many papers on the linear perturbations of a {\it static}
spherical star, notably a series of papers by Thorne and coworkers
{\cite{T1,T2,T3,T4,T5,T6}, another series by Cunningham, Price and
Moncrief \cite{CPM}, and a paper by Ipser and Price
\cite{IpserPrice}. The spherical symmetry allows one to decouple the
perturbations into spherical harmonics.  Because of the
time-independence of the background, one can consider perturbation
modes of the form $\exp(i\omega t)f(r)$ and solve an ODE eigenvalue
problem for the mode functions $f(r)$. The linear perturbations of a
time-{\it dependent} spherical background were evolved in time by
Seidel and coworkers \cite{Seidel}. On a static background two of the
perturbations obey trivial equations, and these have sometimes been
overlooked in counting the degrees of freedom.  The present paper sets
up a mathematical and numerical framework for the study of arbitrary
linear perturbations on a {\it time-dependent} spherically symmetric
background spacetime.

Even if one fixes the coordinate system (also called the gauge) in the
background spacetime, the coordinate freedom of general relativity
poses a fresh problem when linear perturbations are added: one cannot
easily distinguish infinitesimal physical perturbations of the
background from infinitesimal coordinate transformations on the
unchanged background. One can overcome this problem either by fixing
the perturbation gauge, or by introducing linearly gauge-invariant
perturbations. We choose the latter approach because it can easily be
reduced to any particular gauge choice. The construction of the
gauge-invariants, carried out in Sec. III, is a straightforward
application of a general framework (for spherical backgrounds with
arbitrary matter) due to Gerlach and Sengupta (from now on GS)
\cite{GS} that is reviewed in Sec. II.

The equations of GS are covariant in a natural 2+2 split of the
spherically symmetric spacetime. These equations are only ready for
numerical work after we have broken them up into evolution equations
and constraints in Sec. IV. For the axial parity perturbations we can
again use a general prescription due to GS. For the polar
perturbations posing the initial value problem is difficult, and we
use a key idea from Seidel \cite{Seidel} in order to find a subset of
perturbations that can be given free initial data and that evolve
among themselves, thus deserving the name true degrees of
freedom. (The remaining perturbation components are obtained from
these by solving the constraints.)  We use the fact that the fluid
matter provides us with a natural frame field (or set of observers) to
decompose all tensors and tensor equations into scalars and scalar
equations, and to distinguish evolution equations and constraints,
without introducing coordinates.

In many respects our framework is similar to that of Seidel and
coworkers. However, our derivation of the gauge-invariants is more
systematic, and we first make a covariant 2+2 split of the spacetime,
then split the reduced 2-dimensional tensors and tensor equations into
frame components using the natural frame provided by the fluid.  The
final expressions are written so that they clearly display their
causal structure. Being independent of background coordinates and
linearly gauge-invariant, they can easily be specialized to a
particular background coordinate system and perturbation gauge choice,
while going in the other direction would be difficult. Seidel also
restricts to $l=2$ angular dependence, and again it would be difficult
to reconstruct the general from the special case. In the Appendix, we
compare our notation and results to those of Thorne and Seidel.

For completeness, the polar $l=0$ and $l=1$ perturbations, for which
the gauge-invariant treatment breaks down, are discussed here using
equations that are as similar to the ones for $l\ge2$ as possible.  In
previous treatments, with the exception of \cite{T5}, they were often
neglected because they do not couple to gravitational radiation.

Because the main application of this framework is to be stellar
collapse, we must describe the matching of perturbations from the 
interior of the star (with both matter perturbations and gravitational 
waves) to the vacuum exterior (where the matter perturbations are
meaningless). This will be done in a future paper.

In the Appendix, we give the background field equations in
polar-radial and comoving coordinates, discuss the special case of a
static background, and compare our notation with that of Thorne and
coworkers and that of Seidel. Finally, we describe a numerical
algorithm for evolving the perturbations that we have successfully
tested in an application to critical collapse
\cite{critfluid,critscalar}.  It is simple, second order, stable, and
does not require special techniques at the center. As it explicitly
uses the characteristic speeds, it allows us to impose purely outgoing
boundary conditions at the outer boundary of the numerical domain.

\section{Review of the Gerlach and Sengupta framework}

\subsection{The background spacetime}

In describing a spherically symmetric background spacetime and its
linear perturbations we follow the route of reducing the system to 2+2
spacetime dimensions in a covariant manner, without introducing
coordinates. In the following, we use abstract index notation, where
Greek indices denote 4-dimensional spacetime, upper case Latin indices
the 2-dimensional (1+1) reduced spacetime, and lower case Latin
indices the orbits of the spherical symmetry (2-spheres). We write the
general spherically symmetric spacetime as a manifold $M=M^2 \times
S^2$ with metric
\begin{equation}
g_{\mu\nu} \equiv  {\rm diag}\left(g_{AB}, r^2 \gamma_{ab}\right),
\end{equation}
where $g_{AB}$ is an arbitrary Lorentzian metric on $M^2$, $r$ is a
scalar on $M^2$, with $r=0$ defining the boundary of $M^2$, and
$\gamma_{ab}$ is the unit curvature metric on $S^2$. Points in $M^2$
are round spheres of area $4\pi r^2$. (We can use $r$ as a coordinate
on $M^2$, but do not have to.) We introduce covariant derivatives on
spacetime, the subspace $M^2$ and the unit sphere separately, with the
notation
\begin{equation}
g_{\mu\nu;\lambda}\equiv0, 
\qquad g_{AB|C}\equiv0,
\qquad \gamma_{ab:c}\equiv0.
\end{equation}
We shall also need the totally antisymmetric covariant unit tensors on
$M^2$ and $S^2$ respectively:
\begin{equation}
\epsilon_{AB|C}\equiv0, \qquad 
\epsilon_{AC} \epsilon^{BC}\equiv-{g_A}^B,\qquad 
\epsilon_{ab:c}\equiv0,\qquad 
\epsilon_{ac} \epsilon^{bc}\equiv{\gamma_a}^b. 
\end{equation}

We parameterize the stress-energy tensor in
spherical symmetry as
\begin{equation}
t_{\mu\nu} \equiv {\rm diag}\left(t_{AB}, Q r^2 \gamma_{ab}\right).
\end{equation}
 With the shorthand
\begin{equation}
v_A \equiv  {r_{|A}\over r}
\end{equation}
the Einstein equations $G_{\mu\nu}=8\pi t_{\mu\nu}$ in spherical
symmetry, in the 2+2 split, are 
\begin{eqnarray}
\label{G_AB}
-2(v_{A|B} + v_A v_B) 
+ (2{v_C}^{|C} + 3v_C v^C - r^{-2})g_{AB} & = & 8\pi t_{AB} , \\
\label{G4}
{v_C}^{|C} + v_C v^C - {\cal R} & = & 8\pi Q ,
\end{eqnarray}
where ${\cal R}\equiv {1\over 2}{R^A}_A$ is the Gauss curvature of
$g_{AB}$. The equation of stress-energy conservation in spherical
symmetry is
\begin{equation}
{t_{AB}}^{|B}+2 t_{AB}v^B-2v^AQ = 0.
\end{equation}

\subsection{Nonspherical perturbations}

Any linear perturbation around spherical symmetry can be decomposed
into scalar, vector or tensor fields on $M^2$ times spherical
harmonic scalar, vector or tensor fields on $S^2$. The spherical
harmonic scalars on $S^2$ obey
\begin{equation}
\gamma^{ab} {Y_l^m}_{:ab}=-l(l+1)Y_l^m .
\end{equation}
Harmonic vector and tensor fields on $S^2$ can be constructed from the
scalar harmonics. We shall need only vectors and symmetric tensors of
rank two.  A basis of harmonic vector fields on $S^2$ is formed for 
$l\ge 1$ by 
\begin{equation}
{Y_l^m}_{:a}, \qquad {S_l^m}_a\equiv {\epsilon_a}^b {Y_l^m}_{:b}. 
\end{equation}
A basis of harmonic symmetric rank-two tensors is formed by
\begin{equation}
{Y_l^m}\gamma_{ab}, \qquad
{Z_l^m}_{ab}\equiv {Y_l^m}_{:ab}+{l(l+1)\over 2}{Y_l^m}\gamma_{ab}, \qquad
{S_l^m}_{a:b}+{S_l^m}_{b:a}, 
\end{equation}
where the last two expressions vanish identically for $l=0,1$.  Linear
perturbations with different $l,m$ decouple on a spherically symmetric
background. In the following we consider one value of $l,m$ at a time,
and no longer write these indices on $Y$, $S_a$ and
$Z_{ab}$. Furthermore, perturbations with different values of $m$ for
the same $l$ have the same dynamics on a spherically symmetric
background, so that $m$ will never appear in the field equations.
Perturbation fields containing an even power of $\epsilon_{ab}$, such
as $Y_{:a}$, are called polar, or even. They decouple from tensor
perturbations containing an odd power of $\epsilon_{ab}$, such as
$S_a$, which are called axial, or odd. (Note that even and odd in this
sense are not the same as even and odd parity in the standard sense.)

The general axial metric and matter perturbations are parameterized as
\begin{eqnarray}
\Delta g_{\mu\nu} && \equiv   \left(
\begin{array}{cc}
0 &  h_A^{\rm axial} S_a \\
{\rm Symm} &  h (S_{a:b}+S_{b:a})
\end{array}\right), 
\\
\Delta t_{\mu\nu} && \equiv   \left(
\begin{array}{cc}
0 &  \Delta t_A^{\rm axial} S_a\\
{\rm Symm} &  \Delta t(S_{a:b} + S_{b:a})
\end{array}\right).
\end{eqnarray}
The general polar metric and matter perturbations are
\begin{eqnarray}
\Delta g_{\mu\nu} && \equiv   \left(
\begin{array}{cc}
h_{AB}Y &  h_A^{\rm polar} Y_{:a} \\
{\rm Symm} &  r^2 (KY\gamma_{ab}+ GY_{:ab}) 
\end{array}\right),
\\
\Delta t_{\mu\nu} && \equiv   \left(
\begin{array}{cc}
 \Delta t_{AB}Y &   \Delta t_A^{\rm polar} Y_{:a}\\
{\rm Symm} &  r^2 \Delta t^3 Y\gamma_{ab} + \Delta t^2 Z_{ab} 
\end{array}\right).
\end{eqnarray}
We retain the notation of GS, except that we have added the
superscript polar or axial where necessary to remove an
ambiguity. Note also that $r^2$ does not multiply $\Delta t^2$ in the
last equation, and that $\Delta t^3$ and $\Delta t^2$ are scalars.

Let $X$ be an arbitrary tensor field on the background spacetime, and
$\Delta X$ its linear perturbation. Under an infinitesimal coordinate
transformation $x^\mu \to x^\mu + \xi^\mu$, the perturbation is mixed
with the background and transforms as
\begin{equation}
\Delta X \to \Delta X + {\cal L}_{\xi}X.
\end{equation}
The perturbation $\Delta X$ is gauge-invariant to linear order if and
only if ${\cal L}_{\xi}X=0$. The existence of gauge-invariant
perturbations is therefore linked to symmetries of the background
solution.  Because of the spherical symmetry of the background, the
nonspherical perturbations with $l\ge2$ can be made gauge-invariant,
while $l=0$ and $l=1$ perturbations need to be considered separately.

The general infinitesimal coordinate transformation can be
parameterized as 
\begin{equation}
\xi_\mu\equiv\left(\tilde{\xi}_A Y, r^2 \xi Y_{:a} + r^2 M S_{a}\right).
\end{equation}
GS work out how all the bare perturbations defined above
transform. Then they form invariant linear combinations.  Again we
consider polar and axial perturbations separately.  The split is as
follows. There are 3 axial metric perturbations, and 1 axial infinitesimal
coordinate transformation, leaving 2 axial gauge-invariant
perturbations, in the form of a vector field on $M^2$. There are
$7-3=4$ polar gauge-invariant metric perturbations, in the form of a
symmetric tensor and a scalar. There are 3 axial and 7 polar
gauge-invariant matter perturbations.  The axial gauge-invariant
perturbations are
\begin{eqnarray}
l\ge 1: \qquad k_A&\equiv&h_A^{\rm axial}-h_{|A}+2hv_A, \\
L_A&\equiv&\Delta t_A^{\rm axial}-Qh_A^{\rm axial}, \\
l \ge 2: \qquad L&\equiv&\Delta t-Qh,
\end{eqnarray}
and the polar gauge-invariant perturbations are
\begin{eqnarray}
l \ge 0: \qquad k_{AB}&\equiv& h_{AB}-(p_{A|B}+p_{B|A}), \\
k&\equiv&K-2v^Ap_A, \\
T_{AB}&\equiv&
\Delta t_{AB}-t_{AB|C}p^C-t_{AC}p^C_{\ |B}-t_{BC}p^C_{\ |A}, \\
T^3&\equiv&\Delta t^3-(Q_{|C}+2Qv_C)p^C+{l(l+1)\over2}QG, \\
l \ge 1: \qquad T_A&\equiv&
\Delta t_A^{\rm polar}-t_{AC}p^C-\frac{r^2}{2}QG_{|A}, 
\\
l \ge 2: \qquad T^2&\equiv&\Delta t^2-r^2QG, 
\end{eqnarray}
where with GS we define the shorthand
\begin{equation}
p_A\equiv h_A^{\rm polar}-\frac{1}{2}r^2G_{|A}.
\end{equation}
These objects are not gauge-invariant for $l=0,1$, but it is useful to
work with a single set of definitions and field equations for all
$l$. We just have to take into account that for $l=0,1$, $G,h,T^2,L$
vanish and for $l=0$, $h^E_A,h^O_A,T_A,L_A$ also vanish, and we have
to impose an additional gauge fixing in these cases.

To obtain the bare perturbations in an arbitrary gauge, one fixes $h$,
$h_A^{\rm polar}$ and $G$ arbitrarily, and obtains all the other bare
perturbations algebraically from these and the gauge-invariants.

There is also a preferred gauge in which $h=h_A^{\rm polar}=G=0$, the
Regge-Wheeler gauge \cite{RW}. The gauge-invariants have been defined to
correspond one-to-one to the remaining bare perturbations in
Regge-Wheeler gauge (from now on RW gauge). One can therefore describe the GS 
framework also as fixing the gauge to RW gauge. (The GS framework then
still tells us how to transform to any other gauge.) The general axial
metric and matter perturbations, in RW gauge but expressed through the 
gauge-invariants, are
\begin{eqnarray}
\Delta g_{\mu\nu}^{\rm RW} && =  \left(
\begin{array}{cc}
0 &  k_A S_b \\
{\rm Symm} &  0
\end{array}\right), 
\\
\Delta t_{\mu\nu}^{\rm RW} && =  \left(
\begin{array}{cc}
0 &  L_A S_b\\
{\rm Symm} &  L(S_{a:b} + S_{b:a})
\end{array}\right).
\end{eqnarray}
The general polar
metric and matter perturbations are, in RW gauge,
\begin{eqnarray}
\label{evenmetric}
\Delta g_{\mu\nu}^{\rm RW} && =  \left(
\begin{array}{cc}
k_{AB}Y &  0 \\
0 &  kr^2 Y\gamma_{ab} 
\end{array}\right),
\\
\Delta t_{\mu\nu}^{\rm RW} && =  \left(
\begin{array}{cc}
T_{AB}Y &  T_A Y_{:b}\\
{\rm Symm} &  r^2 T^3 Y\gamma_{ab} + T^2 Z_{ab} 
\end{array}\right).
\end{eqnarray}
These expressions are useful both in making calculations and in
interpreting the results.

\section{The perfect fluid}

We now specialize the GS framework to a spherically symmetric perfect
fluid coupled to gravity.  In the perfect fluid approximation the
entropy per particle $s$ is constant along each particle trajectory
(although it may vary between trajectories), and the pressure is
isotropic. That is, we neglect heat fluxes, viscosity and chemical
reactions, and assume that the fluid has only one component. The fluid
is then completely characterized by a single two-parameter equation of
state $p=p(\rho,s)$. Within this class, we allow for arbitrary
equations of state. Often it may be a good approximation to assume the
entropy is constant in space and time, so that the equation of state
is simply $p=p(\rho)$ (barotropic fluid). This case can be recovered
from our results by setting the specific entropy $s$ to a constant and
its perturbation $\sigma$ (introduced below) to zero throughout. A
generalization of our framework to a fluid consisting of several
(noninteracting) components would be straightforward: the relative
abundances of particle types would be treated in the same way as $s$
and $\sigma$.

Instead of giving $p=p(\rho,s)$, many authors give $\rho=\rho(n,s)$
and $p=p(n,s)$ separately, where $n$ is the conserved particle number
density. The two specifications are completely equivalent. We have
chosen the former as it is the simpler one from a spacetime point of
view: $n$ never appears in our equations. 

\subsection{The perfect fluid background spacetime}

The perfect-fluid stress-energy tensor is, 
\begin{equation}
\label{perfectfluid}
t_{\mu\nu} = (p+\rho)u_\mu u_\nu + p g_{\mu\nu},
\end{equation}
where $p$ is the pressure, $\rho$ is the density and $u_\mu$ is the
fluid 4-velocity. In spherical symmetry, $u_\mu = (u_A,0)$.  This
provides an orthonormal basis on $M^2$, namely the timelike unit
vector $u_A$ and the spacelike unit vector
\begin{equation}
n_A \equiv  - \epsilon_{AB} u^B.
\end{equation}
We shall use this basis to transform all tensor fields and tensor
equations on $M^2$ into scalar fields and scalar equations.  The
metric and fundamental antisymmetric tensor on $M^2$ can be written as
\begin{equation}
g_{AB} = - u_A u_B + n_A n_B, 
\qquad \epsilon_{AB} = n_A u_B - u_A n_B.
\end{equation}
For the spherically symmetric background the stress tensor becomes
\begin{equation}
t_{AB} =  \rho \, u_A u_B + p \, n_A n_B, \qquad Q = p
\end{equation}
We  introduce the frame derivatives
\begin{equation}
\label{dotprime}
\dot f \equiv  u^A f_{|A}, \qquad f' \equiv  n^A f_{|A}.
\end{equation}
They will be useful later in discussing the initial value problem.  In
order to write our field equations using only scalar quantities, we
introduce the following background scalars:
\begin{equation}
\label{munu}
\Omega\equiv \ln\rho, \qquad U\equiv u^A v_A=\dot r/r, \qquad W\equiv
n^A v_A = r'/r, 
\qquad \mu\equiv {u^A}_{|A}, \qquad \nu\equiv {n^A}_{|A}.
\end{equation}
Note that $\Omega$, $\mu$ and $U$ are $O(r^0)$ and even as functions
of $r$, with $U-\mu=O(r^2)$, $W$ is odd and $O(r^{-1})$, and $\nu$ is
odd and $O(r)$.

It is useful to define the (spacetime-dependent) Hawking mass $m$ in
spherical symmetry as
\begin{equation}
\label{Hawkingmass}
m\equiv {r\over2}\left(1-r_{,A}r^{,A}\right) 
= {r\over2}\left[1+r^2(U^2-W^2)\right] 
\end{equation}
It is well known that the limit of $m$ in a spacelike and future null
direction gives the ADM and Bondi masses respectively, and that
$m=r/2$ indicates an apparent horizon. $m$ is odd and $O(r^3)$ as
one would expect. It can also be thought of as an integral over the
density, in the sense that
\begin{equation}
 m' = 4\pi r^2 (rW) \rho.
\end{equation}

It is often useful to reparameterize $U$ and $W$ in terms of
\begin{equation}
|v|^2 \equiv v^A v_A = - U^2 + W^2, \qquad V\equiv{U\over W}.
\end{equation}
Note that $|V|<1$, and in fact $V$ is the velocity of the fluid with
respect to constant $r$ observers. $|v|^2$ on the other hand is
related to the Hawking mass via (\ref{Hawkingmass}). 

The frame derivatives obey the following commutation
relation:
\begin{equation}
\label{commutator}
  (\dot f)'-(f')\dot{}=\mu f' - \nu \dot f.
\end{equation}
We also have
\begin{equation}
  u_{A|B}=n_A (n_B \mu - u_B\nu), \qquad
n_{A|B}=u_A(n_B \mu - u_B\nu).
\end{equation}
The fluid equations of motion are given by the conservation of
energy-momentum and by the assumption that the entropy per particle is
constant along particle world lines. In our notation, the matter
equations in spherical symmetry can be written as
\begin{eqnarray}
\label{Omegadot}
\dot\Omega + \left(1+\kay\right)(2U+\mu) & = & 0, \\
\label{Omegaprime}
\kbar \Omega' + \cee s' + \left(1+\kay\right) \nu & = & 0, \\
\label{sdot}
\dot s & = & 0.
\end{eqnarray}
The first of these equations is the energy conservation equation (note
that $\nabla_\mu u^\mu = 2U + \mu$), while the second is the Euler, or
force, equation. The third equation $\dot s=0$ follows from
conservation of the stress-energy tensor (\ref{perfectfluid}) and the
first law of thermodynamics.  Here we have defined the shorthands
\begin{equation}
\kbar\equiv \left({\partial p\over\partial \rho}\right)_s, 
\qquad \cee \equiv {1\over \rho}
\left({\partial p\over\partial s}\right)_\rho.
\end{equation}
We shall confirm later that $\kbar$ is the speed of sound by displaying 
a wave equation for sound waves. 

From the three Einstein equations (\ref{G_AB}) and the identity
(\ref{commutator}) we obtain the relations
\begin{eqnarray}
  U' && = W(\mu - U) 
\label{EinsteinUprime} \\
\dot W && = U(\nu - W) 
\label{EinsteinWdot} \\
W' && = - 4\pi \rho - W^2 + U\mu + {m\over r^3}
\label{EinsteinWprime} \\
\dot U && = - 4\pi p - U^2 + W\nu - {m\over r^3}
\label{EinsteinUdot}
\end{eqnarray}
From the identity ${g^{AB}}_{|AB}=0$, using the fourth Einstein
equation (\ref{G4}) to eliminate ${\cal R}$, we obtain the useful 
relations
\begin{equation}
  \dot\mu - \nu' + \mu^2- \nu^2 = {\cal R} = 
-4\pi\left(p+\rho\right) + {2m\over r^3}.
\end{equation}
We now have a complete list of background identities that we can use
later to simplify the perturbation equations. Expressions for the
scalars in $U$, $W$, $\mu$, $\nu$ and $m$ in specific coordinate
systems are given in the Appendix. Note that $U$ and $\mu$ vanish on a
static background.

\subsection{Nonspherical perfect fluid perturbations}

There should be 5 independent fluid perturbations, namely a density
perturbation, entropy perturbation, 
and a 3-velocity perturbation. It turns out that 4 of these are polar
and 1 is axial.  The polar perturbation of the fluid 4-velocity is
\begin{equation}
\label{Delta_u_ansatz}
\Delta u_{\mu}\equiv \left[\left(\tilde\gamma n_A + {1\over 2}
h_{AB} u^B\right)Y, \
\tilde\alpha Y_{:a}\right],
\end{equation}
while the axial perturbation is just
\begin{equation}
\Delta u_{\mu} \equiv (0,\tilde\beta S_{a}).
\end{equation}
Note that the ansatz for $\Delta u_A$ ensures that
$\Delta(g_{\mu\nu}u^\mu u^\nu)=0$ to linear order.  $\tilde\alpha$ and
$\tilde\beta$ parameterize polar and axial tangential fluid motion, 
while $\tilde\gamma$ parameterizes perturbations of the radial fluid 
motion. The density, pressure, and entropy perturbations are
\begin{equation}
\Delta \rho \equiv \tilde \omega  \rho Y,
\qquad
\Delta s \equiv \tilde \sigma Y,
\qquad
\Delta p = {\partial p\over \partial \rho} \Delta\rho
+ {\partial p\over\partial s} \Delta s = (\kbar \tilde \omega + \cee
\tilde \sigma) \rho Y.
\end{equation}
A gauge-invariant set of fluid perturbations is
\begin{eqnarray}
\label{end_skip}
&&
\alpha \equiv  \tilde \alpha - p^B u_B, 
\qquad \beta \equiv  \tilde \beta, 
\qquad
\gamma \equiv  \tilde \gamma - n^A \left[p^B u_{A|B} + {1\over2} u^B( 
p_{B|A}-p_{A|B})\right], \\
&&
\omega \equiv  \tilde\omega - p^A \Omega_{|A}, \qquad
\sigma \equiv  \tilde\sigma - p^A s_{|A}
\end{eqnarray}
with $p_A$ as defined above.

For the gauge-fixed viewpoint, we give the fluid perturbations in RW
gauge, expressed through the gauge-invariant perturbations. The polar
perturbations are
\begin{equation}
\Delta u_{\mu}^{\rm RW}= \left[(\gamma n_A + {1\over 2} k_{AB} u^B)Y, 
\alpha Y_{:a} \right],
\qquad \Delta \rho^{\rm RW} =  \omega \rho Y,
\qquad \Delta s^{\rm RW} = \sigma Y,
\qquad \Delta p^{\rm RW} =  (\kbar  \omega + \cee \sigma) \rho Y,
\end{equation}
and the axial perturbation is 
\begin{equation}
\Delta u_{\mu}^{\rm RW}=(0,\beta S_a).
\end{equation}
The general gauge-invariant stress-energy perturbations of GS can be
expressed in terms of the gauge-invariant fluid perturbations we have
just defined.  In the axial sector, we have
\begin{equation}
\label{L_A}
L_A =  \beta (\rho + p) u_A, \qquad L=0,
\label{LAform}
\end{equation}
and in the polar sector,
\begin{eqnarray}
&& T_A = \alpha (\rho + p) u_A, 
\qquad T^3 = p k + {\kbar} \rho \omega + \cee \rho \sigma
\qquad T^2 = 0, \\
&& T_{AB} = (\rho + p)\left[\gamma (u_A n_B + n_A u_B) + {1\over 2}
(k_{AC} u_B +
u_A k_{BC}) u^C \right] + \omega  \rho (u_A u_B + \kbar  n_A n_B) 
+ \sigma \cee \rho n_A n_B + p k_{AB}.
\end{eqnarray}
The only axial fluid perturbation, $\beta$, describes equatorial fluid
rotation. $\omega$ describes total density perturbations, $\sigma$
entropy perturbations at constant total density, $\gamma$ describes
radial fluid velocity perturbations, and $\alpha$ describes the
velocity of tangential fluid motion between the poles and the equator
(azimuthal displacement).

\section{The initial value problem for the fluid perturbations}

The previous section has been a straightforward application of the GS
formalism. The perturbed Einstein equations and stress-energy
conservation equations are given in terms of the gauge-invariants and
in 2+2 covariant notation by GS, and we only need to substitute our
parameterization of the gauge-invariant stress-energy perturbations
into their equations. Note that for the perfect fluid the
conservation of stress-energy, plus the conservation of entropy along
particle world lines, is equivalent to the equations of
motion. (For a barotropic fluid with $p=p(\rho)$, stress-energy
conservation alone is sufficient.)
We now turn to the less trivial aspect of breaking these
equations up into evolution equations and constraints, that is, of
posing and physically interpreting an initial value problem. For the
axial sector this is an application of a general method described by
GS. For the polar sector we use an idea due to Seidel \cite{Seidel} to
isolate the true degrees of freedom.

\subsection{$l\ge2$ axial perturbations}

Extracting a well-posed initial value problem from the gauge-invariant
perturbation equations for the axial sector is relatively
straightforward. The one nontrivial matter conservation equation is
(GS4)
\begin{equation}
\label{oddmatter}
(r^2 L_A)^{|A}=(l-1)(l+2)L,
\end{equation}
Note that for perfect fluid matter, we have $L=0$. In our notation,
and using (\ref{Omegadot}) to eliminate $\dot\Omega$, the $\beta$
equation of motion is
\begin{equation}
\label{betadot}
\dot \beta - \kbar (2U+\mu)\beta = 0.
\end{equation}
If the perturbed fluid velocity is to be a regular vector field at the
origin, $\beta$ must scale as $\beta\sim r^{l+1}$. We define
\begin{equation}
\beta\equiv r^{l+1} \bar\beta. 
\end{equation}
The new variable $\bar\beta$ can be
expanded in even positive powers of $r$ near $r=0$. Its equation of
motion is
\begin{equation}
\dot {\bar\beta} + \left[- \kbar (2U+\mu)+(l+1)U\right] \bar\beta = 0.
\end{equation}

As we have $L=0$, Eq. (\ref{oddmatter}) can be integrated to
\begin{equation}
r^2 L_A \equiv \epsilon_{AB} T^{|B} ,
\end{equation}
where $T$ is a scalar which contains all the information about the axial
matter perturbation. The two components of this equation are
\begin{equation}
T'=-r^2(\rho+p)\beta , \qquad \dot{T}=0 .
\end{equation}
[The integrability condition for these equations is just
(\ref{betadot}), using ({\ref{commutator}).]  That is, we set $\beta$
on a given time slice and then we integrate the $T'$ equation for $T$
on that slice. The $\dot{T}=0$ equation states that $T$ is constant
along the integral curves of $u^A$, that is, each particle of the
fluid sees a constant value of $T$ around it. In this sense, the axial
matter perturbation is time-independent, even on a time-dependent
background.

GS have shown that by defining the scalar 
\begin{equation}
\label{Pidef}
\Pi \equiv
\epsilon^{AB}(r^{-2}k_A)_{|B}
\end{equation}
the Einstein equations for $k_A$ can be reduced, for any matter
content, to the scalar wave equation valid for $l\ge 2$
(GS17, corrected GSII6.5b')
\begin{equation} 
\left[\frac{1}{r^2}(r^4\Pi)_{|A}\right]^{|A}-(l-1)(l+2)\Pi
=-16\pi \epsilon^{AB}L_{A|B}
\label{OddMaster}
\end{equation}
which they call the odd-parity master equation. In the case of perfect
fluid matter, the equation of motion of $L_A$ does not contain $\Pi$,
and so can be solved on its own.

The behavior of $k_A$ at $r=0$ can be
calculated by going to Cartesian coordinates. In order to enforce the
correct scaling of $k_A$, we define
\begin{equation}
\Pi\equiv r^{l-2} \bar \Pi,
\end{equation}
where the rescaled field $\bar\Pi$ can be expanded in positive even
powers of $r$ at $r=0$. For $\bar\Pi$ we obtain the field equation
\begin{equation}
\label{Pibar_evolution}
\bar \Pi_{|A}^{\ |A} + 2(l+1)v^A \bar\Pi_{|A}
+(l+2)\left[{{r_{|A}}^{|A}\over r}-(l-1){2m\over r^3}\right] \bar\Pi
= - 16\pi r(p+\rho)
\left\{\bar\beta'+\left[(l+1)W + \frac{\Omega'}{\eor}\right]
\bar\beta\right\}.
\end{equation}
Near the origin, this equation has the approximate form
\begin{equation}
  -\ddot{\bar\Pi}+{\bar\Pi}''+2(l+1)r^{-1}{\bar\Pi}'+ \hbox{explicitly
regular} = 0,
\end{equation}
which is suitable for numerical solution. $\bar\Pi'$ is an odd
function of $r$, so that $\bar\Pi'/r$ is regular at $r=0$.

The scalar $\Pi$ codifies all the information about the evolution of
the axial parity gravitational wave. In fact, once we know $\Pi$ we 
can reconstruct $k_A$ using Eq. (GS16c):
\begin{equation}
(l-1)(l+2) k_A=
16\pi r^2L_A-\epsilon_{AB}(r^4\Pi)^{|B}
=
\epsilon_{AB}\left(16\pi T- r^4\Pi\right)^{|B},
\label{kA_reconstruction2}
\end{equation}
where the last equation holds because of $L=0$.
This equation is valid for $l\ge1$. (For $l=0$ there are no axial
perturbations.) For $l\ge2$, we obtain $k_A$ from $\Pi$. For $l=1$,
however, we obtain a constraint on $\Pi$ instead. The case
$l=1$ has to be treated separately.

The free axial perturbations (for $l\ge2$) are the fluid velocity
perturbation $\beta$, which obeys a transport equation independent of
$\Pi$, and the metric perturbation $\Pi$, which obeys a wave equation
with $\beta$ as a source. On an initial Cauchy surface, one can freely
specify $\beta$, $\Pi$ and $\dot\Pi$, and in this sense there are
three (first-order) degrees of freedom. Note that even though $\beta$
itself cannot oscillate, a non-vanishing $\beta$ can couple to
nonlinear oscillations of the spherical background to generate waves
in $\Pi$.

\subsection{$l=1$ axial perturbations}

Physically, we cannot have any degree of freedom in the metric for
$l=1$ and therefore $\Pi$ cannot obey a wave equation. It has to be
constrained. This is clear in Eq.  (\ref{kA_reconstruction2}), which
for $l=1$ implies:
\begin{equation}
r^4\Pi = 16\pi T + const.
\end{equation}
The integration constant we have introduced here parameterizes a
Kerr-like angular-momentum perturbation that is singular at the
origin. It is zero if we demand a regular center.
Note that even for $l=1$, $\Pi$ is still
gauge-invariant, but $k_A$ cannot be reconstructed uniquely form
it. Now we have to invert Eq.  (\ref{Pidef}), obtaining $k_A$ with a
gauge freedom that is the gradient of an arbitrary scalar.

\subsection{$l\ge2$ polar perturbations}

The polar perturbations are more entangled than the axial ones. Here
we follow Seidel \cite{Seidel} in first focusing attention on those
components of the linear Einstein equations that do not contain matter
variables (except for the entropy perturbation). We shall see that
these Einstein equations (plus the evolution equation for the entropy
perturbation) can be solved autonomously: derivatives of the metric
perturbations contain all the information about the matter
perturbations (except for the entropy perturbation).

We use the fact that the fluid provides a natural frame in $M^2$ to
decompose the symmetric tensor $k_{AB}$ in a coordinate-independent
way into three scalars, as
\begin{equation}
\label{defkAB}
k_{AB} \equiv \eta(-u_A u_B + n_A n_B) + \phi(u_A u_B + n_A n_B) 
            + \psi (u_A n_B + n_A u_B).
\end{equation}
 
The seven linearized Einstein equations of polar parity can be taken
as (GS10b), (GS10c), (GS10d) and (GS12, corrected GSII3.13a). We go
over to a system of scalar equations by substituting the definitions
(\ref{defkAB}), (\ref{dotprime}) and (\ref{munu}) into terms like
$k_{AB|C}$, and projecting the result on the basis $(u,n)$. The
left-hand sides of the perturbed Einstein equations decomposed in this
way are given in \cite{critscalar} (with respect to an arbitrary
frame). Here we give only the components of the right-hand sides for
the perfect fluid (with respect to the fluid frame):
\begin{eqnarray}
l\ge 0: \qquad 
u^A n^B T_{AB} & = & - (\rho+p)\gamma + \frac{1}{2}(\rho-p)\psi, \\
u^A u^B T_{AB} & = & \rho \omega + \rho (\eta-\phi), \\
n^A n^B T_{AB} & = & \rho (\kbar \omega + \cee \sigma) + p (\eta+\phi),
\\
T^3 & = & \rho (\kbar \omega + \cee \sigma) + p k, \\ 
l\ge 1: \qquad u^A T_{A} & = & - (\rho+p) \alpha, \\
n^A T_{A} & = & 0, \\
l\ge 2: \qquad T^2 & = & 0.
\end{eqnarray}
We have indicated for what values of $l$ the corresponding Einstein
equations are valid.  Because $u^A$ is always timelike and $n^A$
always spacelike the scalar decomposition allows us to disentangle
evolution equations and constraints. 

In the remainder of this subsection we discuss the case $l\ge2$, but
we always indicate which equations are also valid for $l=1$ and
$l=0$. The dynamics of $l=1$ and $l=0$ are discussed in the following
two subsections.

For $l\ge 2$ the vanishing of the source $T^2$ implies $\eta=0$.
Among the other six source terms we have just given, there are three
linear combinations that do not contain $\alpha$, $\omega$, or
$\gamma$, and these give rise to three perturbed Einstein equations
not coupled to those three matter perturbations (they do contain the
entropy perturbation $\sigma$). If we choose the three combinations to
be
\begin{eqnarray}
T^3 + 2(n^A T_A)' - n^An^B T_{AB} + 2 (2\nu-W) n^AT_A, \\
\label{soundspeed}
(-\kbar  u^A u^B + n^A n^B) T_{AB} + 4 W n^AT_A, \\
n^A T_A, 
\end{eqnarray}
and introduce the new variable
\begin{equation}
\chi\equiv\phi - k + \eta,
\end{equation}
which from now on replaces $\phi$ as an independent variable, then the
corresponding Einstein equations take the form
\begin{eqnarray}
l\ge 1: \qquad
\label{chi_free}
 - \ddot \chi + \chi''  + 2 ( \mu - U )\psi' && =  S_\chi, \\
\label{k_free}
 - \ddot k +  \kbar k''  -2 c_s^2 U\psi'  && = S_k, \\
\label{psi_free}
 -\dot \psi  && = S_\psi.
\end{eqnarray}
These are the core equations of our paper.  Only the highest
derivatives of all variables, which determine the causal structure of
the equations, have been written out explicitly.  The source terms
$S_\chi$, $S_k$ and $S_\psi$ are homogeneously linear in $\chi$ and
$k$, and their first derivatives $\chi'$, $k'$, $\dot\chi$, and $\dot
k$, as well as $\psi$ and $\sigma$ (undifferentiated), with
background-dependent coefficients, but do not contain any higher
derivatives. They are given in Appendix \ref{full_equations}.

We choose the remaining three components of the Einstein equations as
\begin{eqnarray}
(u^A n^B + n^A u^B)T_{AB}, \\
u^Au^BT_{AB}, \\
u^AT_A,
\end{eqnarray}
which give us
\begin{eqnarray}
l\ge 0: \qquad
\label{gamma_constraint}
8 \pi(\rho+p) \gamma & = & (\dot k)' + C_\gamma, \\
\label{omega_constraint}
8 \pi \rho \omega & = & - k'' + 2U\psi' + C_\omega, \\
l\ge 1: \qquad
\label{alpha_constraint}
16\pi(\rho+p) \alpha & = & \psi' + C_\alpha, 
\end{eqnarray}
where again all the highest derivatives have been written out
explicitly, and the right-hand sides do not contain $\alpha$, $\gamma$
or $\omega$, (or indeed $\sigma$). These equations give $\alpha$,
$\gamma$ and $\omega$ as spatial derivatives of the metric
perturbations.  Seen the other way around, they form an ODE system for
$\psi$, $k$ and $\dot k$ in the radial coordinate which can be solved
from the center outwards, with boundary conditions at the center given
by local flatness. The correspondence between $\alpha$, $\gamma$ and
$\omega$ on the one hand, and $\psi$, $k$ and $\dot k$ on the other
(for given data $\chi$ and $\dot\chi$, and $\sigma$) is therefore
one-to-one. (With one proviso discussed at the end of this section.)

The linearized stress-energy conservation
equation (GS15a,b, corrected GSII3.14a,b) are equivalent to the
following equations of motion for the matter perturbations.
\begin{eqnarray}
l\ge 0: \qquad
\label{omegadot}
-\dot\omega-\left(1+\kay\right)\left(\gamma+\frac{\psi}{2}\right)' 
& = & S_\omega, \\
\label{gammadot}
\left(1+\kay\right)\left(\gamma-\frac{\psi}{2}\right)\dot{} + 
\kbar\omega' 
& = & S_\gamma, \\
l\ge 1: \qquad
\label{alphadot}
-\dot\alpha & = & S_\alpha.
\end{eqnarray}
Again we have written out only the highest derivatives.  The three
source terms are linear in the (undifferentiated) matter perturbations
$\alpha$, $\gamma$ and $\omega$, and $\sigma$, as well as $\chi$, $k$,
$\chi'$, $k'$, $\dot\chi$, $\dot k$, and $\psi$. The full equations
are given in Appendix \ref{full_equations}.  It is possible to remove
$\dot\psi$ and $\psi'$ from (\ref{omegadot}) and (\ref{gammadot})
using the Einstein equations (\ref{psi_free}) and
(\ref{alpha_constraint}).  The matter perturbation equations then do
not contain highest derivatives of the metric perturbations. In this
form they are most amenable to numerical solution.  We obtain
\begin{eqnarray}
l\ge 1:
\label{evol_omega}
-\dot\omega-\left(1+\kay\right)\gamma'
& = & \bar S_\omega, \\
\label{evol_gamma}
\left(1+\kay\right)\dot\gamma + \kbar\omega' 
& = & \bar S_\gamma.
\end{eqnarray}
To these matter equations we have to add the perturbation of the
trivial entropy equation $\dot s=0$, which is
\begin{equation}
l\ge 0: \qquad \dot\sigma+\left(\gamma+\frac{\psi}{2}\right)s'=0.
\label{sigmadot}
\end{equation}

We now introduce auxiliary variables $\chi',k',\dot\chi,\dot k$ in
order to consider the perturbation equations as a system of {\it
first-order} system of evolution equations and constraints for the
variables
\begin{equation}
u\equiv\{\chi,\psi,k,\alpha,\gamma,\omega,\sigma,\chi',k',\dot\chi,\dot k\}.
\end{equation}
We consider an equation that contains $\dot u$ an evolution equation,
and an equation that contains only $u$ and $u'$ a
constraint. Constraints can be solved within a single timelike
hypersurface. (If $u^A$ is not normal to the surface of our choice, we
need to use the evolution equations in order to write the constraints
as ODEs in the radial coordinate.)  Apart from the equations we have
already given explicitly, the first-order form introduces trivial
evolution equations for $\chi$ and $k$, and trivial constraints for
$\chi'$ and $k'$. Altogether we have 11 evolution equations and 5
constraints for 11 first-order variables $u$. This means that there
are, in a first-order in time sense, 6 true degrees of freedom, or 6
functions of the radial coordinate that can be specified freely on a
Cauchy surface. A natural choice of these would be the matter
perturbations $\alpha$, $\gamma$, $\omega$ and $\sigma$, plus $\chi$
and $\dot\chi$ which describe polar gravitational waves. [There are
two ``polarizations'' of gravitational waves, namely axial ($\Pi$) and
polar ($\chi$).]  The metric perturbations $\psi$, $k$ and $\dot k$
would then be obtained from the constraints. While this scheme is the
most natural in terms of a split into matter motion and gravitational
waves, it maximizes the number of constraints that must be
solved. This is an inconvenience both in terms of numerical work and
numerical stability.

Seidel \cite{Seidel} has noticed that one can identify the 6 true
degrees of freedom with the initial data for the metric perturbations
alone, plus the entropy perturbation $\sigma$. This means that
\begin{equation}
u_{\rm free}\equiv\{\chi,\psi,k,\sigma,\dot\chi,\dot k\}
\end{equation}
can be set freely and evolve among themselves. In a first-order
formulation, there remain the trivial two constraints and two
evolution equations for the auxiliary variables $\chi'$, $k'$,
$\dot\chi$ and $\dot k$, but the three nontrivial constraints are now
only used if one wants to reconstruct $\alpha$, $\gamma$ and $\omega$
as spatial derivatives of the free Cauchy data $u_{\rm free}$.

Let us now consider the causal structure of the equations from a
spacetime point of view.  The highest derivatives of $\chi$ form a
wave equation with characteristics given by the metric $g_{\mu\nu}$,
and as we have seen, initial data $\chi$ and $\dot\chi$ can be set
independently from the matter perturbations. Therefore, $\chi$ can
reasonably be said to parameterize gravitational waves inside the
matter.  The two first-order evolution equations for $\omega$ and
$\gamma$ are equivalent to either of the two second-order equations
\begin{eqnarray}
- \ddot \gamma + \kbar \gamma'' + ... && = 0, \\ 
- \ddot \omega + \kbar\omega'' + ... && = 0,
\end{eqnarray}
where we have only written out the highest derivatives.  These are
wave equations with characteristics given by the ``fluid metric''
\begin{equation}
 g_{\mu\nu}^{\rm fluid}\equiv{\rm diag}\left(-\kbar u_Au_B+ n_An_B,
r^2 \gamma_{ab}\right)
\end{equation}
The characteristics have speed $c_s$ relative to the fluid background. 
The evolution equation for $k$ is a wave equation with the same
characteristics. It ``knows about'' the speed of sound because $\kbar
$ appears in the linear combination, Eq. (\ref{soundspeed}), of
Einstein equations that gives rise to this equation. It is surprising
that the metric perturbation $k$ describes sound waves. The analogy is
correct, nevertheless, as the data $\omega$ and $\gamma$ are
essentially the spatial derivatives of the data $k$ and $\dot k$. It
may be helpful to think of $k$ as describing ``longitudinal
gravitational waves'', made physical by the presence of matter.
Finally, $\psi$ is advected with the fluid, as its evolution equation
contains $\dot\psi$ but not $\psi'$. The same is true for the matter
perturbation $\alpha$, and not surprisingly, $\alpha$ is essentially
the spatial derivative of $\psi$. $\alpha$ describes tangential fluid
motion between the poles and the equator, and is therefore coupled to
the density perturbation $\omega$ that acts as a restoring force.

Finally, we consider the behavior of the polar perturbations at the
origin.  We require that the metric perturbation (\ref{evenmetric}) be
a regular tensor in four dimensions at $r=0$, as defined by going to
Cartesian, Minkowski-like coordinates near $r=0$. We find that any
regular solution must scale at $r=0$ as
\begin{equation}
\label{regularity}
l\ge 0: \qquad
\chi \equiv  r^{l+2}\bar\chi, \qquad 
\psi \equiv r^{l+1}\bar\psi , \qquad 
k \equiv r^l\bar k, \qquad 
\eta \equiv r^l \bar \eta,
\end{equation}
where the barred variables can all be expanded in even positive powers
of $r$. 
This behavior at the origin is consistent with the equations
of motion, in the sense that Eqs. (\ref{chi_free}),
(\ref{k_free}) and (\ref{psi_free}) can be solved for $\ddot{\bar\chi}$,
$\ddot{\bar k}$, and $\dot{\bar \psi}$ order by order in $r$, for
arbitrary values of $(\dot{\bar\chi},\bar\chi,\dot{\bar k},\bar
k,\bar\psi,\sigma)$.  

To investigate the regularity of
the velocity perturbations at the origin, we work in RW gauge, and
introduce Minkowski-like coordinates $t,x,y,z$ near the center. For
simplicity of presentation we only consider the case $m=0$. The
four-velocity perturbation becomes
\begin{eqnarray}
\nonumber
\Delta u_\mu^{\rm RW} dx^\mu = &&
-\left[\left(\eta-{\phi\over2}\right)Y_{l0}\right]dt
-\left[\alpha r^{-1}{1\over\sin\theta} {dY_{l0}\over d\theta}\right] dz
\\
&& 
+ \left[\left(\gamma-{\psi\over2}\right)r^{-1}Y_{l0}+\alpha
r^{-2}{\cos\theta\over\sin\theta} {dY_{l0}\over d\theta}\right]
(xdx+ydy+zdz) 
\end{eqnarray}
This is regular if and only if the terms in square brackets are
regular scalars. Now $\cot\theta\
dY_{l0}/d\theta=-lY_{l0}+O[(\cos\theta)^{l-2}]$ and so
the leading orders of $\alpha$ and $\gamma$ can both be a factor of $r^2$
lower than naively expected if they are correlated. Taking this
possibility into account, the matter
perturbations near the origin behave as
\begin{equation}
l\ge 1: \qquad
\alpha  =  r^l \bar\alpha, \qquad 
\gamma = r^{l-1} \bar\gamma, \qquad 
    \text{with }\bar\gamma=l\bar\alpha+O(r^2) \text{ \ at $r=0$}, \qquad
\omega = r^l \bar\omega, \qquad \sigma=r^l \bar\sigma.
\end{equation}
To avoid the constraint between the leading orders of $\bar\gamma$ and
$\bar\alpha$, one could replace $\gamma$ by $\gamma-\alpha'$ as a
dependent variable that would be unconstrained, similar to the
variable $\chi$. The present notation, however, gives rise to the more
transparent evolution equations.  Again, the scaling at the origin is
consistent with the equations of motion (\ref{omegadot}),
(\ref{gammadot}) and (\ref{alphadot}). In particular, what looks like
a leading term that is too large by a factor of $r^{-2}$ cancels in
each case. Furthermore, the constraint between the leading terms of
$\alpha$ and $\gamma$ is conserved by the evolution.

The constraint equations (\ref{gamma_constraint}),
(\ref{omega_constraint}), (\ref{alpha_constraint}) can be solved
consistently for $\bar\gamma$, $\bar\omega$ and $\bar\alpha$, order by
order in $r$. The converse is almost true. For given $\bar\chi$,
$\dot{\bar\chi}$, $\bar\sigma$, $\bar\alpha$, $\bar\omega$ and
$\bar\gamma$, these equations can be solved order by order for
$\bar\psi$, $\bar k$ and $\dot{\bar k}$, with the exception of the
leading order $\bar k_0$ of $\bar k$ and the leading order
$\bar\psi_0$ of $\bar\psi$. The gravitational wave degrees of freedom
are therefore the functions $\bar\chi$, $\dot{\bar \chi}$, plus the
two numbers $\bar k_0$ and $\bar\psi_0$, while the matter degrees of
freedom are either the functions $\bar\omega$, $\bar\gamma$,
$\bar\alpha$, or alternatively the functions $\bar k$ without $\bar
k_0$, $\dot{\bar k}$, and $\bar\psi$ without $\bar\psi_0$. In practice
there is no need to distinguish between matter and metric
perturbations. 

\subsection{$l=1$ polar perturbations}

The case of $l=1$ for the polar perturbations differs from $l\ge2$ in
two important aspects. First, the trivial field equation that stated
that $\eta=0$ no longer holds, so we have an additional
variable. Second, the variables we have defined for $l\ge2$ exist for
$l=1$, but are only partially gauge-invariant.  Under a gauge
transformation generated by the vector $\xi_\mu dx^\mu = \tilde{\xi}_A
Y dx^A + r^2 \xi Y_{:a} dx^a$, we have the following change in the
metric perturbations:
\begin{eqnarray}
k_{AB} &\longrightarrow& 
k_{AB} + (r^2\xi_{|A})_{|B} + (r^2\xi_{|B})_{|A} , \\
k &\longrightarrow& k + 2\xi + (r^2)^{|A} \xi_{|A} .
\end{eqnarray}
Decomposed in the fluid basis, this is:
\begin{eqnarray}
\label{etagauge}
\eta &\longrightarrow& \eta + 2 r^2 (-U\dot\xi+W\xi')
+ r^2(-\ddot\xi+\xi''+\nu\xi'-\mu\dot\xi) , \\
\label{chigauge}
\chi &\longrightarrow& \chi 
- 2\xi + 2r^2 \left[ W\xi'+\xi''+(U-\mu)\dot\xi \right], \\
\label{psigauge}
\psi &\longrightarrow& \psi -2 r^2 (U\xi'+W\dot\xi)
-r^2 \left[ (\dot\xi)'+(\xi')\dot{}-\mu\xi'-\nu\dot\xi \right] , \\
\label{kgauge}
k &\longrightarrow& k + 2\xi + 2 r^2(-U\dot\xi+W\xi') .
\end{eqnarray}
We need to impose one gauge condition in order to fix the gauge
freedom parameterized by $\xi$. It is tempting to extend Seidel's free
evolution scheme to $l=1$. Clearly, one can use the gauge freedom to
set $\eta=0$ by solving (\ref{etagauge}) as a wave equation for
$\xi$. This still leaves us with two free functions to specify the
gauge completely, such as $\xi$ and $\dot\xi$ on one spacelike
slice. As there are no dipole ($l=1$) gravitational waves, one would
also like to make the gauge choice $\chi=0$, as we have seen that
$\chi$ obeys a wave equation at the speed of light. While the two free
functions could be used to set $\chi$ and $\dot\chi$ to zero on one
slice, $\chi$ could not be made to disappear on every
slice. Conversely, one might be able to solve (\ref{chigauge}) as a
second-order constraint equation for $\xi$ on each slice in order set
$\chi=0$, which would leave us with one free function of one variable,
such as $\eta$ at the center as a function of time. ($\eta'$ at the
center would be fixed by imposing regularity). That means that we
would be stuck with at least this constraint to solve. Furthermore,
the presence of the term $(U-\mu)\dot\xi$ in (\ref{chigauge})
complicates the interpretation as a constraint, as the hypersurface on
which this equation becomes an ODE in the radial coordinate is not a
natural one to choose for any other purpose.

All this is not promising. Instead we follow Thorne \cite{T5} in
choosing the gauge $k=0$. We might call this gauge the radial
perturbation gauge, as the background scalar $r$ remains the area
radius in the perturbed spacetime. We now use the matter equations
(\ref{alphadot})--(\ref{sigmadot}) to evolve the matter 
perturbations $\omega$, $\gamma$, $\alpha$ and $\sigma$. 
These evolution equations contain $\eta$, $\psi$ and
$\chi$ (but not their derivatives), while $k$ vanishes by our gauge
choice. We can solve the five Einstein equations (\ref{k_free}),
(\ref{psi_free}), (\ref{gamma_constraint}), (\ref{omega_constraint}),
(\ref{alpha_constraint}) for $\dot\chi$ and $\chi'$, $\dot\psi$ and
$\psi'$, but only for one linear combination of $\dot\eta$ and
$\eta'$, namely $D\eta$, where the differential operator $D$ is
defined as
\begin{equation}
Df = r^{-1}|v|^{-2}(Wf'-U\dot f) = {r^{|A} f_{|A}\over r^{|B}r_{|B}}
\end{equation}
which is just $\partial/\partial r$ on a spacelike surface that is
everywhere orthogonal to the surfaces $r={\rm const.}$ (a polar
slice). The expression for $D\eta$ is given in Appendix 
\ref{full_equations}. The
remaining sixth Einstein equation (\ref{chi_free}) does not tell us
anything new: as it contains $\ddot\chi$, it follows as a
consequence of the other equations. 
The fact that we do not have
$\dot\eta$ and $\eta'$ independently means that we can integrate the
constraint for $\eta$ only on a surface that is normal everywhere to
the surfaces $r={\rm const.}$ As we are restricted to these surfaces
anyway, we also give only $D\chi$ and $D\psi$ in Appendix 
\ref{full_equations}.

The three constraint equations form three coupled first-order linear
ODEs. To analyze them we define again regularized variables:
\begin{eqnarray}
\chi\equiv r^3\bar\chi , \qquad
\psi\equiv r^2\bar\psi , \qquad
\eta\equiv r  \bar\eta , \\
\gamma\equiv  \bar\gamma, \qquad
\alpha\equiv r\bar\alpha, \qquad
    \text{with }\bar\gamma=\bar\alpha+O(r^2) \text{ \ at $r=0$}, \qquad
\omega\equiv r\bar\omega, \qquad
\sigma\equiv r\bar\sigma.
\end{eqnarray} 
The constraints given in Appendix \ref{full_equations} give two 
relations among the leading terms of these new variables:
\begin{eqnarray}
\bar\chi &=& \frac{8\pi\rho}{5}
\left[ \bar\omega - 8 U \bar\alpha \eor \right] 
- 2 U^2 \bar\eta + O(r^2)
\text{ \ \ at $r=0$} , \\
\bar\psi &=& 8\pi \bar\alpha (\rho+p) + 2 U \bar\eta + O(r^2)
\text{ \ \ at $r=0$} .
\end{eqnarray}

Note that imposing $k=0$ we still have the residual freedom of
functions $\xi$ obeying $\xi + r^3 |v|^2 D\xi = 0$. We have to give a
boundary condition for this ODE at $r=0$ at each instant of time, and
therefore the residual freedom is a single function of time. We can
use this function to set $\bar\eta=O(r^2)$ at the center.

\subsection{$l=0$ polar (spherical) perturbations}

The case of spherical ($l=0$ polar) perturbations has the same
problem with $\eta$ as the $l=1$ case. It differs in that the polar
matter perturbation parameterized by $\alpha$ does not exist, and that
there are now two (polar) gauge degrees of freedom, with no remnants of
gauge-invariance.  We impose the gauge
\begin{equation}
k=0, \qquad \hat\psi\equiv\psi-\frac{2UW}{U^2+W^2}(\eta-\chi)=0.
\end{equation}
Here $\hat\psi$ is $\psi$ in the radial (instead of the fluid) frame,
see Appendix \ref{polar_radial}. In this gauge we have two 
metric perturbations, $\eta$ and $\chi$, which obey constraints given 
in Appendix \ref{full_equations}. As in the
$l=1$ case, only the combination $D\eta$ of $\dot\eta$ and $\eta'$ is
known.  Using these constraints, the metric perturbations can be
calculated from the matter perturbations $\gamma,\omega,\sigma$. 
This gauge is the only one in which we have found simple evolution
equations. Note that a natural coordinate choice for the background
spacetime are polar-radial coordinates which are defined by
$g_{\theta\theta}=r^2$, $g_{tr}=0$. $D$ is then just
$\partial/\partial r$ in these coordinates, and the gauge choice
$k=\hat\psi=0$ just means that the perturbed metric (which is still
spherical) has the same form as the background metric. Here we give
the spherical perturbation equations in our notation, rather than in
polar-radial coordinates, for the purpose of a unified presentation of
all perturbations. Given that the spherical perturbations are so much
messier than the $l\ge2$ perturbations, however, one might as well
calculate them in the same gauge choice one has already adopted for
the background spacetime.

The matter perturbations obey the evolution equations 
(\ref{omegadot}), (\ref{gammadot}), (\ref{sigmadot}), which contain 
derivatives of the metric perturbations. The latter can be eliminated, 
using the perturbed Einstein equations. The resulting equations are
given in Appendix \ref{full_equations}.

Again, we introduce regularized variables:
\begin{eqnarray}
\chi\equiv r^2\bar\chi , \qquad
\eta\equiv    \bar\eta , \\
\gamma\equiv  r \bar\gamma, \qquad
\omega\equiv    \bar\omega, \qquad
\sigma\equiv    \bar\sigma .
\end{eqnarray} 
Note that now we do not have any cancellation of leading orders in
$r$, so we obtain the powers of $r$ naively expected. The constraints
give only one condition at the center:
\begin{equation}
\bar\chi=\frac{8\pi\rho}{3}\bar\omega - 2 U^2 \bar\eta + O(r^2)
\text{ \ \ at $r=0$.}
\end{equation}
Again, we have a residual gauge freedom that we can use to set 
$\bar\eta=O(r^2)$ at the center.

\section{Conclusions}

We have given the field equations for all spherical and non-spherical
linear perturbations of a spherically symmetric but time-dependent
self-gravitating perfect fluid. In this task we have applied a
formalism created by Gerlach and Sengupta \cite{GS}. We have
generalized and clarified previous results by Thorne and coworkers
\cite{T1,T2,T3,T4,T5,T6}, Ipser and Price \cite{IpserPrice} and Seidel
\cite{Seidel}. Our formulation is distinguished by the following nice
features.

1) Our perturbation variables are linearly gauge-independent (with the
exception of the polar $l=0,1$ perturbations, see below). This means
that we can be sure to only count physical degrees of freedom. Our
results can be translated into any particular gauge (for example
Regge-Wheeler gauge) by explicit algebraic formulas.

2) Using the fact that the fluid provides a preferred frame, we have
decomposed both tensor variables and tensor equations into frame
components. This procedure clarifies how many degrees of freedom are
present, which perturbed Einstein equations are evolution equations
and which are constraints, and what are the characteristic speeds. We
find that in the sense of free functions on a Cauchy surface, the
axial $l=1$ perturbations have 1 gauge-invariant degree of freedom,
the axial $l\ge2$ perturbations have 3 (including one gravitational
wave polarization), the polar $l=0$ perturbations 3, the polar $l=1$
perturbations 4, and the polar $l\ge2$ perturbations 6 (including the
other gravitational wave polarization and sound waves).

3) All the final background and perturbation equations are given in
terms of frame derivatives of spacetime scalars or frame components of
tensors. This means that our equations can be easily translated into
any coordinate system for the background (for example comoving or
polar-radial coordinates).

4) With the exception of the polar $l=0,1$ perturbations (see below),
we have split the gauge-invariant perturbations into a set of
variables that can be specified freely on a Cauchy surface, and that
evolve freely among themselves, and another set of variables that are
determined by the first set through algebraic constraints. 
Writing out only 
the highest derivatives of each variable, the axial perturbation
free evolution equations for $l\ge2$ are
\begin{eqnarray}
\nonumber
 - \ddot \Pi + \Pi''   && =  \dots, \\
\nonumber
\dot \beta  && = \dots,
\end{eqnarray}
and the polar free evolution equations for $l\ge2$ are
\begin{eqnarray}
\nonumber
 - \ddot \chi + \chi''  + 2 ( \mu - U )\psi' && =  \dots, \\
\nonumber
 - \ddot k +  \kbar k''  -2 c_s^2 U\psi'  && = \dots , \\
\nonumber
\dot \psi  && = \dots, \\
\nonumber
\dot \sigma && = \dots,
\end{eqnarray}
where $c_s^2$ is the local sound speed, and the coefficients $U$ and
$\mu$ vanish in a static background.  Having a set of free evolution
equations that are manifestly hyperbolic is helpful for numerical
work, in particular if one wants to use characteristic methods. (An
example is given in the Appendix.)

5) No gauge-invariant variables exist for the polar $l=0,1$
perturbations. Nevertheless we define the same variables for $l=0,1$
as for $l\ge2$. As these variables are not fully gauge-invariant for
$l=0,1$, additional constraints must be imposed to fix the gauge, but
we can make use of the fact that our variables are still partially
gauge-invariant for $l=1$. This clarifies the role of the polar $l=1$
perturbations, and in particular the number of degrees of freedom.

6) We allow for a two-parameter equation of state $p=p(\rho,s)$ and
take into account perturbations of the entropy $s$. Our framework can
easily be generalized to multi-component fluids. 

We believe that our formulation characterizes the perturbation initial
value problem more clearly and concisely than previous work, and is
also a good starting point for its numerical solution.

\acknowledgments

We would like to thank Jes\'us Ma\'\i z, Vincent Moncrief, Ed Seidel
and Bernd Schmidt for helpful conversations, and Toni Font, John
Friedman and Kip Thorne for helpful comments on the manuscript.
J.M.M. would like to thank the Albert-Einstein-Institut for
hospitality.

\appendix

\section{Full polar perturbation equations}
\label{full_equations}

Here are the missing source terms in
Eqs. (\ref{chi_free})--(\ref{psi_free}) and
(\ref{gamma_constraint})--(\ref{evol_gamma}). Recall that some of the equations
to which these sources belong are not valid for $l=0$ or $l=1$.  On
the other hand, note that $\eta=0$ for $l\ge 2$. Note also that we
have used the background equations to eliminate all occurrences of
$\dot\Omega$, $\Omega'$, $\dot U$, $U'$, $\dot W$, $W'$, and $\dot \mu
- \nu'$.

\begin{eqnarray}
\label{S_chi}
\nonumber
S_\chi  & = &  
- 2 \left[ 2 \nu^2 + 8 \pi \rho - {6m\over r^3} 
         - 2 U ( \mu -U )
    \right] ( \chi + k )  
+ \frac{(l-1)(l+2)}{r^2} \chi
\\ \nonumber &&
+ 3 \mu \dot\chi
+ 4 ( \mu - U ) \dot k
- ( 5 \nu - 2 W ) \chi'
- 2 [ 2 \mu \nu - 2 ( \mu - U ) W + \mu' - \dot\nu ] \psi
\\ &&
+2 \eta''
- 2 ( \mu - U ) \dot\eta
+ ( 8 \nu - 6 W ) \eta'
- \left[ - 4 \nu^2 + \frac{l(l+1)+8}{r^2} + 8 \nu W 
         + 4 ( 2\mu U + U^2 - 4W^2 - 8 \pi \rho )
  \right] \eta,
\\
\label{S_k}
\nonumber S_k
& = &
  ( 1 + c_s^2 ) U \dot\chi
+ [ 4 U + c_s^2 ( \mu + 2 U ) ] \dot k
- W ( 1 - c_s^2 ) \chi'
- ( \nu + 2 W c_s^2 ) k'
\\ \nonumber &&
- \left[ 2 \left( \frac{1}{r^2} - W^2 \right) + 8 \pi p 
       - c_s^2 \left( \frac{l(l+1)}{r^2} + 2 U ( 2\mu + U ) - 8 \pi \rho
               \right)
  \right] ( \chi + k )
\\ \nonumber &&
- \frac{(l-1)(l+2)}{2 r^2} ( 1 + c_s^2) \chi
+ 2 [ - \mu W ( 1 - c_s^2 ) + ( \nu + W ) U ( 1 + c_s^2 ) ] \psi
+ 8\pi \cee \rho \sigma
\\ &&
- 2 U \dot\eta + 2 W \eta'
+ \left[ \frac{l(l+1)+2}{r^2} - 6 W^2 + 16 \pi p - 2 U (2\mu+U) c_s^2
  \right] \eta,
\\
S_\psi
& = &
2 \nu ( \chi + k ) + 2 \mu \psi + \chi' -2 \eta (\nu-W) - 2 \eta',
\\
C_\gamma
& = &
- W \dot\chi + U \chi' - ( \mu - 2 U ) k'
+ \frac{1}{2} \left[ \frac{l(l+1)+2}{r^2} + 2 U ( 2 \mu + U )
                     - 2 W ( 2 \nu +  W ) + 8 \pi (p-\rho)
  \right] \psi
- 2 U \eta',
\\
C_\omega
\nonumber & = &
  \left[ \frac{l(l+1)}{r^2} + 2 U ( 2\mu + U ) - 8 \pi \rho
  \right] ( \chi + k )
- \frac{(l-1)(l+2)}{2 r^2} \chi
+ 2 [ \nu U + ( \mu + U ) W ] \psi
\\ &&
+ U \dot\chi + ( \mu + 2 U ) \dot k
+ W \chi' - 2 W k'
- 2 \eta U ( 2 \mu + U ),
\\
C_\alpha
& = &
2 \mu ( \chi + k ) + 2 \nu \psi + \dot\chi + 2 \dot k 
- 2 \eta ( \mu + U ),
\\
S_\omega
& = &
\left( 1 + \frac{p}{\rho} \right)
\left[ - \frac{l(l+1)}{r^2} \alpha
       + \frac{\dot\chi + 3\dot k}{2}
       + \left( \nu + 2 W - \frac{\nu}{c_s^2} \right) 
         \left( \gamma + \frac{\psi}{2} \right)
\right] 
+ ( \mu + 2 U ) \left( c_s^2 - \frac{p}{\rho} \right) \omega 
\nonumber \\ &&
- \cee \left[ \left(\gamma+\frac{\psi}{2}\right)\frac{s'}{c_s^2}
           - \sigma (\mu+2U) 
    \right],
\\
S_\gamma
& = &
\left( 1 + \frac{p}{\rho} \right)
\left[ \frac{\chi' + k' - 2 \eta'}{2}
       + \left[ c_s^2 (\mu + 2 U ) - \mu \right]
         \left( \gamma - \frac{\psi}{2} \right) 
\right]
- \nu \left( c_s^2 - \frac{p}{\rho} - \frac{\rho + p}{c_s^2}
               \frac{\partial c_s^2}{\partial \rho}
      \right) \omega
\nonumber \\ &&
- \cee \sigma'
-\sigma\left[ \cee \left(\nu-\frac{s'}{c_s^2}
                         \frac{\partial c_s^2}{\partial s}\right)
            + s' \frac{\partial \cee}{\partial s}
            - \nu \left(1+\frac{p}{\rho}\right) 
              \frac{1}{c_s^2}\frac{\partial c_s^2}{\partial s}
       \right]
- \omega s' \left[ \frac{\partial c_s^2}{\partial s}
                 - \cee \left(1+\frac{\rho}{c_s^2}
                   \frac{\partial c_s^2}{\partial\rho}\right)
            \right],
\\
S_\alpha
& = &
- \frac{k+\chi}{2} + \eta - c_s^2 ( \mu + 2 U ) \alpha 
+ \frac{c_s^2 \omega + \cee \sigma}{1+\frac{p}{\rho}}, \\
\nonumber
\bar S_\omega & = & 
\left( 1 + \frac{p}{\rho} \right)
\left[ \left( - \frac{l(l+1)}{r^2} + 8\pi(\rho+p) \right) \alpha
       + \frac{\dot k}{2}
       + (\mu+U) \eta - \mu (\chi+k)
\right] 
+ ( \mu + 2 U ) \left( c_s^2 - \frac{p}{\rho} \right) \omega 
\\ && 
+ \cee ( \mu+2U ) \sigma
- \frac{1}{c_s^2} 
  \left[ s' \cee + \left(1+{p\over\rho}\right) ( \nu-2W\kbar ) \right]
  \left( \gamma+\frac{\psi}{2} \right) 
+ \nu \left(1+{p\over\rho}\right) \left( \gamma-\frac{\psi}{2} \right),
\\
\nonumber
\bar S_\gamma & = &
\left( 1 + \frac{p}{\rho} \right)
\left[ \frac{k'}{2}
       + \left( c_s^2 (\mu + 2 U ) - \mu \right)
         \left( \gamma - \frac{\psi}{2} \right) 
       - \mu \psi - \nu (\chi+k) + (\nu-W) \eta
\right]
\\ \nonumber &&
- \cee \sigma'
- \sigma \cee \left[ \nu 
                   + \frac{s'}{\cee} \frac{\partial\cee}{\partial s}
                   - \left( \frac{\nu}{\cee} 
                            \left( 1 + \frac{p}{\rho} \right)
                          + s'
                     \right) 
                     \frac{1}{c_s^2} \frac{\partial c_s^2}{\partial s}
              \right]
\\ &&
+ \omega 
  \left[ \nu \left(\frac{p}{\rho}-c_s^2\right)
       + s' \left( \cee - \frac{\partial c_s^2}{\partial s} \right)
       + \left[ \nu (\rho+p)+ \rho \cee s' \right] 
         \frac{1}{c_s^2} \frac{\partial c_s^2}{\partial\rho}
  \right] .
\end{eqnarray}

The matter perturbation equations for the polar $l=1$ perturbations are
the same as for $l\ge2$, namely (\ref{evol_omega}), (\ref{sigmadot}). 
The three constraint equations for the polar $l=1$ perturbations that
we require are
\begin{eqnarray}
\nonumber
r |v|^2D\eta &=&
  4\pi\rho(1-\kbar)\omega-16\pi(\rho+p)U\alpha
- 4\pi \rho \cee \sigma
\label{Deta}
\\ &&
- \left({2\over r^2}-3W^2+8\pi p+U^2\right)\eta
- \left[W^2+U^2-4\pi(\rho+p)\right]\chi
- 2UW\psi , \\
\label{Dchi}
r|v|^2D\chi & = & 8\pi\rho\omega-32\pi(\rho+p)U\alpha
-\left({2\over r^2}+2U^2-8\pi\rho\right)\chi
+2\left[U\nu-(\mu+U)W\right]\psi-2U^2\eta, \\
r|v|^2D\psi & = & 8\pi(\rho+p)(\gamma+2W\alpha)
+2(\mu W-\nu U)(\eta-\chi)+4UW\eta
-\left[U^2-W^2+{2\over r^2}+4\pi(p-\rho)\right]\psi
\label{Dpsi}.
\end{eqnarray}

The matter evolution equations for the polar $l=0$, or spherical
perturbations are 
\begin{eqnarray}
-\dot\omega - \eor \gamma' 
&=&
  (\mu+2U) \cee \sigma
- \omega \left[ 4\pi \frac{U}{|v|^2} (\rho+p)
              + (\mu+2U) \left( \frac{p}{\rho} - c_s^2 \right) \right]
\nonumber \\ &&
- \gamma \left[ \frac{\cee s'}{c_s^2}
              - \eor \left( -4\pi \frac{W}{|v|^2} (\rho+p)
                          + \nu + 2W - \frac{\nu}{c_s^2} \right)
         \right]
\nonumber \\ &&
- ( \chi - \eta )
  \left[ -\frac{UW}{U^2+W^2} \frac{\cee s'}{c_s^2} \right.
\nonumber \\ &&
       \left. + \eor \left( \mu + U 
                   - \nu \frac{UW}{U^2+W^2} \frac{1+c_s^2}{c_s^2}
                   + \frac{U}{2} \frac{|v|^2}{U^2+W^2}
                   + 4\pi \frac{U W^2}{-U^4+W^4} (\rho+p) \right)
  \right]
\nonumber \\ &&
- \eta \frac{U}{|v|^2} \eor \left(-\frac{1}{2r^2} + 4\pi \rho\right)
, \\
\eor \dot\gamma + c_s^2 \omega' 
&=&
  \gamma \eor \left( -4\pi \frac{U}{|v|^2} (\rho+p)
                   - \mu + (\mu+2U) c_s^2 \right)
\nonumber \\ &&
- \sigma \left[ \nu \left( \cee - \eor \frac{1}{c_s^2} 
                                  \frac{\partial c_s^2}{\partial s}
                    \right)
              + 4\pi \frac{W\cee}{|v|^2} (\rho+p)
              + s' \left( \frac{\partial\cee}{\partial s}
                        - \frac{\cee}{c_s^2}
                          \frac{\partial c_s^2}{\partial s} \right)
         \right]
\nonumber \\ &&
- \omega \left[ \nu \left( c_s^2 - \frac{p}{\rho} 
                         - (\rho+p) \frac{1}{c_s^2}
                           \frac{\partial c_s^2}{\partial\rho} \right)
              + 4\pi \frac{W c_s^2}{|v|^2} (\rho+p)
              - s' \left( \cee - \frac{\partial c_s^2}{\partial s}
                        + \frac{\rho \cee}{c_s^2}
                          \frac{\partial c_s^2}{\partial\rho} \right)
         \right]
\nonumber \\ &&
- (\chi-\eta) \eor
  \left( \nu - \mu \frac{UW}{U^2+W^2}(1+c_s^2)
       - \frac{2U^2W}{U^2+W^2} c_s^2
       + \frac{W}{2} \frac{|v|^2}{U^2+W^2}
       + 4\pi \frac{U^2 W}{U^4-W^4} (\rho+p)
  \right)
\nonumber \\ &&
- \eta \frac{W}{|v|^2} \eor \left( \frac{1}{2r^2} + 4\pi p \right) 
- \cee \sigma', \\
\dot{\sigma}&=&-s'\left(\gamma+\frac{UW}{U^2+W^2}(\eta-\chi)\right).
\end{eqnarray}
The remaining two metric perturbations $\eta$ and $\chi$ can be
obtained from the constraints
\begin{eqnarray}
r|v|^2D\eta &=& 
  4\pi(\rho+p) \left( \chi + \frac{2U^2}{|v|^2} \eta \right)
+ 8\pi (\rho+p) \frac{2UW}{|v|^2} \gamma
+ 4\pi \rho \frac{U^2+W^2}{|v|^2} (\cee\sigma + (1+c_s^2)\omega)
, \\
\nonumber
r|v|^2D\chi &=& 
\frac{4UW}{U^2+W^2}
\left( \mu W - \nu U + 4\pi \frac{UW}{|v|^2} (\rho+p) \right) 
( \chi - \eta )
+ \left( -\frac{1}{r^2} + 8\pi \rho \right)
  \left( \chi + \frac{2U^2}{|v|^2} \eta \right)
\\ &&
+ 8\pi (\rho+p) \frac{2UW}{|v|^2} \gamma
+ 8\pi \rho \frac{U^2+W^2}{|v|^2} \omega.
\end{eqnarray}

\section{Background expressions in polar-radial coordinates}
\label{polar_radial}

Excepting null coordinates, there are two natural coordinate systems
for spherically symmetric fluid spacetimes. Polar radial coordinates
$r,t$ use the scalar $r$ as one coordinate, and make $t$ orthogonal to
it, so that the metric is
\begin{equation}
ds^2 = - \alpha^2 dt^2 + a^2 dr^2 + r^2 d\Omega^2
\label{Schw_coordinates}
\end{equation}
The gauge is fixed completely when $\alpha$ is fixed to be 1 at $r=0$
(or at $r=\infty$).
In these coordinates we have
\begin{eqnarray}
U = {u^r\over r}, \qquad W={n^r\over r}, \qquad |v|^2 = {1\over
a^2r^2}, 
\qquad {2m\over r^3}=1-a^{-2}\\
u^t={1\over \alpha\sqrt{1-V^2}},  \qquad u^r={V\over a\sqrt{1-V^2}}, \\
n^t={V\over \alpha\sqrt{1-V^2}},  \qquad n^r={1\over a\sqrt{1-V^2}}, \\
\mu=u^t\left({a_{,t}\over a}+{VV_{,t}\over 1-V^2}\right)+
n^r\left({V\alpha_{,r}\over \alpha}+{V_{,r}\over 1-V^2}\right), \\
\nu=u^t\left({Va_{,t}\over a}+{V_{,t}\over 1-V^2}\right)+
n^r\left({\alpha_{,r}\over \alpha}+{VV_{,r}\over 1-V^2}\right).
\end{eqnarray}
The first derivatives of the metric that appear here are given by the
Einstein equations (\ref{G_AB}),
\begin{eqnarray}
\frac{a^2 - 1}{2r} + {a,r\over a} &=& 4\pi a^2r{\rho+V^2 p\over 1-V^2} 
, \\
\frac{1 - a^2}{2r} + {\alpha,r\over \alpha} &=& 4\pi a^2r{V^2\rho+p\over 1-V^2}
, \\
-{a,t\over \alpha} &=& 4\pi a^2r{V(\rho+p)\over 1-V^2} .
\end{eqnarray}
The Einstein equation (\ref{G4}) is a combination of derivatives of
these three. The matter evolution equations are, eliminating all
derivatives of the metric variables:

\begin{eqnarray}
(1-\kbar V^2)\frac{\rho_{,t}}{\alpha}+(1-\kbar)V\frac{\rho_{,r}}{a}&=&
\frac{1}{a} \left[ 4\pi a^2 r (\rho+p)^2 V
                      - (\rho+p)(\frac{2V}{r}+V_{,r})
                      + V \cee\rho s_{,r}(1-V^2)
                 \right] , \\
(1-\kbar V^2)\frac{V_{,t}}{\alpha}+(1-\kbar)V \frac{V_{,r}}{a}&=&
\frac{1}{a} (1-V^2)
\left[ -4\pi a^2 r (\rho \kbar V^2+p)
     - (1-V^2) \frac{\rho\cee s_{,r}+\kbar \rho_{,r}}{\rho+p}
     + \frac{1-a^2}{2r}
     + \frac{3+a^2}{2r}\kbar V^2
\right] , \\
\frac{s_{,t}}{\alpha}+V \frac{s_{,r}}{a}&=& 0 .
\end{eqnarray}

\section{Background expressions in comoving coordinates, and
comparison with Seidel}

The metric in comoving coordinates, using the notation of Seidel
\cite{Seidel}, is
\begin{equation}
ds^2 \equiv - N^2 d\tau^2 + A^2 d\mu^2 + R^2 d\Omega^2.
\end{equation}
Note there are three independent metric functions, and that Seidel's
$R$ is our $r$. The condition that lines of constant $\mu$ are particle
worldlines leads to
\begin{equation}
u^\tau=N^{-1},\quad u^\mu=0, \quad n^\tau=0, \quad n^\mu=A^{-1}.
\end{equation}
This, and fixing the value of $N$ at the center fixes the coordinate
freedom, up to a relabeling of the radial coordinate $\mu\to
\mu'(\mu)$.  In these coordinates we have (using our notation on the
left-hand side, and Seidel's on the right-hand side)
\begin{equation}
U = {R_{,\tau}\over NR}, \qquad W = {R_{,\mu}\over AR}, \qquad
\mu = {A_{,\tau}\over AN}, \qquad \nu = {N_{,\mu}\over AN},
\qquad {2m\over r^3}=1-{R_{,\mu}^2\over A^2}+{R_{,\tau}^2\over N^2}.
\end{equation}
Seidel fixes the gauge freedom in $\mu$ by imposing $A=(4\pi\rho
R^2)^{-1}$. This means that his radial coordinate is mass-like, with
$\mu\simeq 4\pi r^3\rho/3$ to leading order.

The gauge-invariant metric perturbations of Seidel \cite{Seidel} are
related to our variables in the following way.  (Seidel's notation is
used on the left-hand sides, and our notation on the right-hand
sides.)
\begin{eqnarray}
q_1 & = & \phi+\eta, \\
q_2 & = & k,\\
N^{-1} \pi_1 & = & (\phi+\eta)\dot{}+2\psi'+2\nu\psi+\mu(\phi-\eta), \\
N^{-1} \pi_2 & = & \dot k + 2W\psi +U(\phi-\eta), \\
N^{-1}A^{-1}R^2\pi_4 & = & \psi.
\end{eqnarray}
The $\eta$ terms are contained in Seidel's definition, but $\eta=0$
for $l\ge2$ by one of the Einstein equations. Seidel's conditions for
regularity at the origin for $l=2$, although somewhat complicated in
appearance, are equivalent to our conditions (\ref{regularity}) if one
takes into account Seidel's' definition of $A$ and $\mu$.

\section{Static background, and comparison with Thorne {\it et al.}}
\label{static_bg}

As previous work by Thorne and coworkers, and other authors, was
restricted to finding pulsation modes of a static background solution,
we should point out here how their results fit into our wider
framework. 

On a static
background the fluid and radial bases coincide, and therefore the
comoving and radial coordinates coincide. We choose a coordinate
system of the form $-N^2dt^2+A^2dr^2+r^2d\Omega^2$.  (Thorne uses
$N^2=e^\nu$ and $A^2=e^\lambda$.)  The background scalars are:
\begin{equation}
U=\mu=0, \qquad W=\frac{1}{rA} , \qquad 
\nu=Ar\left(\frac{m}{r^3}+4\pi p\right) .
\end{equation}
$\dot\nu$ and all other time derivatives also vanish. 

Our axial gauge-invariant perturbations, when restricted to RW
gauge, are related to the RW gauge axial perturbations of Thorne by
\begin{equation}u^A k_A = \frac{1}{N}h_0, \qquad
   n^A k_A = \frac{1}{A}h_1, \qquad
   \beta = \frac{1}{N} U_{{\rm T},t} ,
\end{equation}
where we have introduced the subscript $T$ to distinguish Thorne's
variable from our $U$ background scalar.

We first consider the $l\ge 2$ axial perturbations. On a static
background we have $\dot\beta=0$, and this agrees with equation
$U_{T,tt}=0$ in Appendix B of \cite{T1}.  Ref. \cite{T1} also states
that axial-parity gravitational waves do not couple to the stellar
matter, because the only matter perturbations that have wave-like
solutions (namely sound waves) are polar. This is true only on a
static background: Eq.  (\ref{Pibar_evolution}) shows that the axial
gravitational waves can couple to a non-vanishing $\beta$ times an
oscillating background coefficient.

For the axial $l=1$ perturbations, we still have $\dot\beta=0$, and
$\beta$ is the only physical perturbation, corresponding to the
``$\Omega$'' of \cite{T5}. 

We now consider the $l\ge 2$ polar perturbations. For a static
background, $\mu=U=0$, and therefore $\psi$ and $\psi'$ are not
present in the wave equations for $\chi$ and $k$ [see
Eqs. (\ref{S_chi})--(\ref{S_k})].  Furthermore, as $S_\psi$ does not
contain $\psi$ or $\psi'$, $\psi$ can be obtained simply by
integrating $-\dot\psi=S_\psi=2\nu(\chi+k)+\chi'$.  In this sense,
there are only 4 dynamical degrees of freedom (not counting the
entropy perturbation), while $\psi$ plays a passive role.
Nevertheless, $\psi$ can be specified freely in the initial data, and
in this sense there are really 5 degrees of freedom (not counting the
entropy perturbation), even on the static background.

Let us see how this fits into Thorne's formalism.
Our metric gauge-invariant metric perturbations, when restricted to RW
gauge, are related to the RW gauge metric perturbations of Thorne by
\begin{equation}
\label{Thornevariables}
\eta=\frac{H_0-H_2}{2} , \qquad
\phi=-\frac{H_2+H_0}{2} , \qquad
\psi=\frac{H_1}{N A} , \qquad
k=-K .
\end{equation}

Our matter perturbations $\alpha,\gamma,\omega$ are related to those of
Thorne, $V_{\rm T},W_{\rm T}$, by
\begin{equation}
\gamma+\frac{\psi}{2}=-\frac{1}{r^2 N}W_{{\rm T},t} , \qquad
\alpha=\frac{1}{N}V_{{\rm T},t} , \qquad
\omega=\eor 
\left[
- \frac{1}{r^2A}
      \left( W_{{\rm T},r} + W_{\rm T} \frac{\rho_{,r}}{\rho+p} \right)
- \frac{l(l+1)}{r^2} V_{\rm T} + K + \frac{H_2}{2}.
\right] 
\end{equation}
Again, we have added the subscript T to distinguish two of Thorne's
variables which might be confused with our variables.  We have
re-obtained all equations (8a-d) ,(9a-c), (C3-4), (C6-7) in \cite{T1}
[except for a wrong $r^2$ factor in (9a)].  It is implicit in these
equations that $V_{\rm T}$, $V_{{\rm T},t}$, $W_{\rm T}$, $W_{{\rm
T},t}$, $K$ and $K_{,t}$ can all be set freely on an initial Cauchy
surface. Note however that $V_{\rm T}$ and $W_{\rm T}$
(undifferentiated) only enter into the equations through their
combination in $\omega$, and we consider only this combination as
physical. Thus we count 5 polar perturbation degrees of freedom.

In discussing the degrees of freedom in more detail we first neglect
entropy. The fundamental fluid perturbation variable in the framework
of Thorne and coworkers is a 3-dimensional displacement vector
parameterized by $U_{\rm T}$ (axial) and $V_{\rm T}$ and $W_{\rm T}$
(polar). Its time derivative constitutes the fluid velocity
perturbation, corresponding to our variables $\beta$, $\alpha$ and
$\gamma$.  The displacement generates a perturbation in the
particle-number density (and hence in the energy density and
pressure).

In our framework, we take the point of view that apart from the
velocity perturbations, only the density perturbation $\omega$ (a
linear combination of $V_{\rm T}$ and $W_{\rm T}$) is observable, but
not the fluid displacements $U_{\rm T}$, $V_{\rm T}$ and $W_{\rm T}$
separately. In other words, two perfect fluid initial data sets in
which the fluid particles are arranged differently, but with the same
(Eulerian) density and velocity, are considered experimentally
indistinguishable.

The framework of Thorne and coworkers has therefore two degrees of
freedom more than ours. They are $U_{\rm T}$ (axial), and a linear
combination of $V_{\rm T}$ and $W_{\rm T}$ (polar) that is linearly
independent of $\omega$ (for example just $V_{\rm T}$). These degrees
of freedom play only a passive role dynamically, and would be
considered "gauge" in our framework.  In Thorne's approach they are
considered physical, with the assumption that there exist other
properties that provide physical "labels" on the particles.

Finally, we come back to entropy. In our formalism we explicitly add
an Eulerian perturbation $\sigma$ to the degrees of freedom to account
for any kind of entropy perturbation. If the entropy is not constant
on the background solution, a displacement of the particles would
generically produce a (Laplacian) entropy perturbation. The two linear
combination of $V_{\rm T}$ and $W_{\rm T}$ could then be interpreted
as the density and entropy perturbations, and both would be physical.
(But entropy is not considered in \cite{T1,T2,T3,T4,T5,T6}, and so this
is our speculation on an extension of that framework.)

For the polar $l=1$ perturbations, we count 3 degrees of freedom,
namely $\alpha$, $\omega$ and $\gamma$, while \cite{T5} implies that 4
functions can be specified freely in the initial data, namely $V_{\rm
T}$, $V_{{\rm T},t}$, $W_{\rm T}$ and $W_{{\rm T},t}$. This
discrepancy is the same as in the general $l\ge2$ case.

Ipser and Price \cite{IpserPrice} have made a Fourier ansatz for the
$l\ge2$ polar perturbations of a static star, with a barotropic
equation of state $p=p(\rho)$ for both the background and
perturbations, thus addressing the same problem as Thorne and
coworkers. They find a system of two coupled second-order ODEs for the
variables $H_0$ and $K$ of (\ref{Thornevariables}). This is similar to
the equations we have obtained for $\chi$ and $k$, minus the equation
for $\psi$ (which plays only a passive role on the static background.)

\section{Change of frame}
\label{change_frame}

Given the fluid basis $\{u_A,n_A\}$, any other orthonormal basis
$\{\ur_A,\nr_A\}$ can be described as a hyperbolic rotation (boost)
through an angle $\xi(x^D)$ with respect to the fluid basis:
\begin{equation}
\left(\matrix{u_A \cr n_A}\right)
=
\left(\matrix{\cosh\xi & \sinh\xi \cr \sinh\xi & \cosh\xi}\right)
\left(\matrix{\ur_A \cr \nr_A}\right)
\equiv 
R(\xi)
\left(\matrix{\ur_A \cr \nr_A}\right)
\end{equation}

This rotation $R(\xi)$ induces a transformation in those
scalar-variables which are basis-dependent. First, if we decompose
$k_{AB}$ in the same way as Eq. (\ref{defkAB}) but now using the new
basis, we have the same $\eta$, but $\phi$ and $\psi$ change as
\begin{equation}
\left(\matrix{\phir \cr \psir}\right)
=
R(2\xi)
\left(\matrix{\phi \cr \psi}\right)
\end{equation}
Note that the perturbations $\alpha,\beta,\gamma,\omega$ are
basis-independent by definition.

The new frame derivatives acting on a scalar $f$ are related to the
old derivatives by a single rotation:
\begin{equation}
\left(\matrix{ \dr{f} \equiv \ur^A f_{|A} \cr 
               \pr{f} \equiv \nr^A f_{|A}    } \right)
=
R(-\xi)
\left(\matrix{\dot f \cr f^\prime}\right).
\end{equation}

We can change to a new set of projections of $v^A$ on the new basis
and new Ricci-rotation coefficients $\mur,\nur$:
\begin{equation}
\left(\matrix{ \Ur \equiv \ur^A v_{A} \cr 
               \Wr \equiv \nr^A v_{A}    } \right)
=
R(-\xi)
\left(\matrix{U \cr W}\right),
\qquad
\left(\matrix{ \mur \equiv \ur^A{}_{|A} \cr 
               \nur \equiv \nr^A{}_{|A}    } \right)
=
R(-\xi)
\left(\matrix{\mu \cr \nu}\right)
-\left(\matrix{\pr{\xi} \cr \dr{\xi}}\right)
=
R(-\xi)
\left(\matrix{\mu-\xi' \cr \nu-\dot\xi}\right)
\end{equation}

Of particular interest besides the fluid basis is the radial basis,
where $\nr^A=v^A/v$ and $\ur^A=-\epsilon^{AB}\nr_A$. This basis is
described by $\xi={\rm arcth}(V)$, that is, both bases are related
by the boost of velocity $V$.
The equations of motion for spherical perturbations take their
simplest form in the radial base, not the fluid base. In particular,
it is useful to impose the gauge $\hat\psi=0$. 

\section{Aspects of a numerical implementation}
\label{numerical}

In a previous application of the framework given here
\cite{critfluid,critscalar} we have used a numerical evolution scheme
that explicitly uses the characteristic speeds and is second order in
space and time. The set of free evolution equations is of the form
\begin{equation}
{\partial u\over \partial t} + A(r,t) {\partial u \over \partial r} 
= B(r,t) u
\end{equation}
in a time coordinate $t$ and a radial coordinate $r$, where $A$ and
$B$ are matrices of rank 3 for the axial perturbations and of rank 6
for the polar perturbations. Let $V$ be the matrix of (column)
eigenvectors of $A$. Let $\Lambda$ be the diagonal matrix composed of
the corresponding eigenvalues. Then $A=V \Lambda V^{-1}$. Let
$\Lambda_+$ be $\Lambda$ with zeros in the place of the negative
eigenvalues, and $\Lambda_-$ be $\Lambda$ with zeros in the place of
the positive eigenvalues. Let $A_+=V \Lambda_+ V^{-1}$, and let $A_-=V
\Lambda_- V^{-1}$, so that $A=A_++A_-$. $A$ is quite sparse,
and its eigenvalues and eigenvectors can be calculated in closed
form. With this splitting of $A$ given, we use a finite differencing
scheme that is a type of iterated Crank-Nicholson scheme, while using
the information about left-moving and right-moving modes. We iterate
the equation
\begin{equation} 
u^{n+1}_j =  u^n_j + \Delta t \left[ (A_-)^{n+{1\over2}}_j
{4u^{n+{1\over2}}_{j+1} - u^{n+{1\over2}}_{j+2}- 3 u^{n+{1\over2}}_j
\over2 \Delta r} - (A_+)^{n+{1\over2}}_j {4u^{n+{1\over2}}_{j-1} -
u^{n+{1\over2}}_{j-2}- 3 u^{n+{1\over2}}_j \over2 \Delta r} +
B^{n+{1\over2}}_j u^{n+{1\over2}}_j \right].
\end{equation}
Here the auxiliary variable $u^{n+{1\over2}}$ is estimated as $u^n$
for the first, forward-in-time, step, and then corrected on subsequent
steps as $u^{n+{1\over2}}=(u^n+u^{n+1})/2$. 

The boundary $r=0$ does not require special treatment, as $u(-r)=\pm
u(r)$ for all $u$, so that ghost grid points with $r<0$ are available
for taking derivatives. The one-sided differencing operators we use
here do not give exactly zero at $r=0$ even if analytically $\partial
u/\partial r(0)=0$, but that is consistent: all terms in the finite
difference equations combine so that numerically, too, $u(0)=0$ at all
times if $u(0)=0$ initially. This also ensures that source terms of
the form $u/r$ in the evolution equations are well behaved
numerically, and do not pose any stability problems.



\end{document}